\documentclass[10pt,twoside,twocolumn,aps,prl,superscriptaddress,notitlepage,nobibnotes]{revtex4-2}
\usepackage[utf8]{inputenc}
\usepackage[letterpaper]{geometry}
\usepackage{color}
\usepackage{textcomp}
\usepackage{amsmath}
\usepackage{amssymb}
\usepackage{graphicx}
\usepackage[unicode=true,pdfusetitle,
 bookmarks=false,
 breaklinks=true,pdfborder={0 0 1},backref=false,colorlinks=true]
 {hyperref}
\hypersetup{
 pdfborderstyle=,citecolor=blue,linkcolor=black,urlcolor=blue}

\makeatletter
\usepackage{color}
\usepackage{xcolor}\usepackage{soul}\usepackage{multirow}\usepackage{verbatim}
\usepackage{amsfonts}\usepackage[calcwidth,explicit]{titlesec}
\usepackage{chngcntr}
\usepackage[normalem]{ulem}

\renewcommand{\frontmatter@abstractwidth}{\dimexpr\textwidth-3cm\relax}

\titleformat{\section}{}{}{0pt}{}
\titleformat{\section}{\bfseries\sffamily\large\filcenter}{\thesection.}{0.2em}{#1}
\titlespacing{\section}{0pt}{0.2ex}{0.2ex}
\titleformat{\paragraph}[runin]{\normalfont\normalsize\bfseries}{}{0pt}{\theparagraph.}
\titlespacing*{\paragraph}{0em}{0ex}{0.3em}[]

{}

\setcitestyle{super}

\renewcommand{\thesection}{\Roman{section}}
\counterwithout{paragraph}{subsubsection}
\renewcommand{\theparagraph}{\arabic{paragraph}}
\def\p@paragraph{}

\AtBeginDocument{
\renewcommand{\ref}[1]{\autoref{#1}}
\renewcommand{\figureautorefname}{Fig.}\renewcommand{\equationautorefname}{Eq.}
\renewcommand{\tableautorefname}{Tbl.}
\renewcommand{\sectionautorefname}{\S}\renewcommand{\subsectionautorefname}{\S}
\renewcommand{\paragraphautorefname}{\S}
}

\makeatother

\begin{document}
\setlength{\abovedisplayskip}{0.2ex}\setlength{\belowdisplayskip}{0.2ex} 
\setlength{\abovedisplayshortskip}{0.2ex}\setlength{\belowdisplayshortskip}{0.2ex} 
\title{Targeted Writing and Deleting of Magnetic Skyrmions \\
in Two-Terminal Nanowire Devices}
\author{Soong-Guen Je}
\affiliation{Center for X-ray Optics, Lawrence Berkeley National Laboratory, Berkeley,
California 94720, USA}
\affiliation{Department of Emerging Materials Science, Daegu Gyeongbuk Institute
of Science and Technology (DGIST), Daegu 42988, Korea}
\affiliation{Department of Physics, Chonnam National University, Gwangju 61186,
Korea}
\author{Dickson Thian}
\affiliation{Institute of Materials Research \& Engineering, Agency for Science,
Technology \& Research (A{*}STAR), 138634 Singapore}
\author{Xiaoye Chen}
\affiliation{Institute of Materials Research \& Engineering, Agency for Science,
Technology \& Research (A{*}STAR), 138634 Singapore}
\affiliation{Data Storage Institute, Agency for Science, Technology \& Research
(A{*}STAR), 138634 Singapore}
\author{Lisen Huang}
\affiliation{Institute of Materials Research \& Engineering, Agency for Science,
Technology \& Research (A{*}STAR), 138634 Singapore}
\affiliation{Data Storage Institute, Agency for Science, Technology \& Research
(A{*}STAR), 138634 Singapore}
\author{Dae-Han Jung}
\affiliation{School of Materials Science and Engineering, Ulsan National Institute
of Science and Technology, Ulsan 44919, Korea}
\author{Weilun Chao}
\affiliation{Center for X-ray Optics, Lawrence Berkeley National Laboratory, Berkeley,
California 94720, USA}
\author{Ki-Suk Lee}
\affiliation{School of Materials Science and Engineering, Ulsan National Institute
of Science and Technology, Ulsan 44919, Korea}
\author{Jung-Il Hong}
\affiliation{Department of Emerging Materials Science, Daegu Gyeongbuk Institute
of Science and Technology (DGIST), Daegu 42988, Korea}
\author{Anjan Soumyanarayanan}
\email{anjan@imre.a-star.edu.sg}

\affiliation{Institute of Materials Research \& Engineering, Agency for Science,
Technology \& Research (A{*}STAR), 138634 Singapore}
\affiliation{Data Storage Institute, Agency for Science, Technology \& Research
(A{*}STAR), 138634 Singapore}
\affiliation{Physics Department, National University of Singapore (NUS), 117551
Singapore}
\author{Mi-Young Im}
\email{mim@lbl.gov}

\affiliation{Center for X-ray Optics, Lawrence Berkeley National Laboratory, Berkeley,
California 94720, USA}
\affiliation{Department of Emerging Materials Science, Daegu Gyeongbuk Institute
of Science and Technology (DGIST), Daegu 42988, Korea}
\begin{abstract}
\noindent Controllable writing and deleting of nanoscale magnetic
skyrmions are key requirements for their use as information carriers
for next-generation memory and computing technologies. While several
schemes have been proposed, they require complex fabrication techniques
or precisely tailored electrical inputs, which limits their long-term
scalability. Here we demonstrate an alternative approach for writing
and deleting skyrmions using conventional electrical pulses within
a simple, two-terminal wire geometry. X-ray microscopy experiments
and micromagnetic simulations establish the observed skyrmion creation
and annihilation as arising from Joule heating and Oersted field effects
of the current pulses, respectively. The unique characteristics of
these writing and deleting schemes, such as spatial and temporal selectivity,
together with the simplicity of the 2-terminal device architecture,
provide a flexible and scalable route to the viable applications of
skyrmions.
\end{abstract}
\maketitle

\section*{Introduction\label{sec:intro}}

Magnetic skyrmions are spatially localized spin textures with a well-defined
topology \citep{Nagaosa.2013,Soumyanarayanan.2016j7e}. Their nanoscale
size and ambient stability in metallic thin films \citep{Soumyanarayanan.2017,Boulle.2016,Moreau-Luchaire.2016,Je.2020},
as well as efficient coupling to electrical currents \citep{Jiang.2015,Woo.2016},
are desirable inherent attributes that have prompted much excitement
\citep{Fert.2017}. Notably, numerous recent device proposals seek
to harness the skyrmion motion within a wire geometry \citep{Parkin.2008}
to realize applications in memory, logic, and unconventional computing
\citep{Kang.2016,Pinna.2018,Prychynenko.2018}. An indispensable requirement
for such devices is a spatially and temporally controlled scheme to
write and delete skyrmions through robust and scalable techniques.

There has been considerable progress in writing magnetic skyrmions.
Theoretical studies suggested that skyrmions can be created by spin
currents \citep{Iwasaki.2013,Lin.20135iv,Lin.2013p,Sampaio.2013}.
Recent device-level write efforts have mostly utilized spin-orbit
torques (SOTs) in application-relevant materials \citep{Fert.2017,Miron.2011,Iwasaki.20135wa,Sampaio.2013}.
To facilitate the SOT-writing, the schemes require additional components,
for example, geometric constrictions \citep{Jiang.2015,Hrabec.2017,Finizio.2019},
and defects \citep{Buttner.2017,Woo.2018}. While effective, however,
these additional requirements lead to fabrication complexities in
some cases, limiting their scalability.

On the contrary, the electrical deletion of skyrmions has thus far
been lacking, and has been recognized as a critical pending challenge
toward functional skyrmionic devices \citep{Buttner.2017,Woo.2018}.
While two recent schemes have been reported for deterministic deletion
and creation of skyrmions \citep{Finizio.2019,Woo.2018}, they have
additional prerequisites, such as in-plane bias field \citep{Woo.2018}
or precisely positioned nanostructures \citep{Finizio.2019}. In the
other, completely different current paths are required for writing
and deletion \citep{Finizio.2019}, increasing the complexity of device
operation. Therefore, the need for the hour is to realize simple schemes
for electrical skyrmion writing and deletion that may be readily adapted
to myriad syrmion applications in a controlled fashion.

Here we present a simple and facile approach for writing and deleting
skyrmions wherein an identical conventional current pulse is used
for both operations. Crucial to this achievement is the harnessing
of two by-products of electrical currents: Joule heating and Oersted
field. By exploiting contrasting spatial and temporal properties of
these two effects, we can achieve targeted operation, e.g., by simply
reversing the pulse polarity. The ubiquitous character of these current-induced
effects across material systems, coupled with minimalist requirements,
offers much-needed attributes of scalability and broad applicability.
\begin{figure*}
\centering\includegraphics[width=0.8\textwidth]{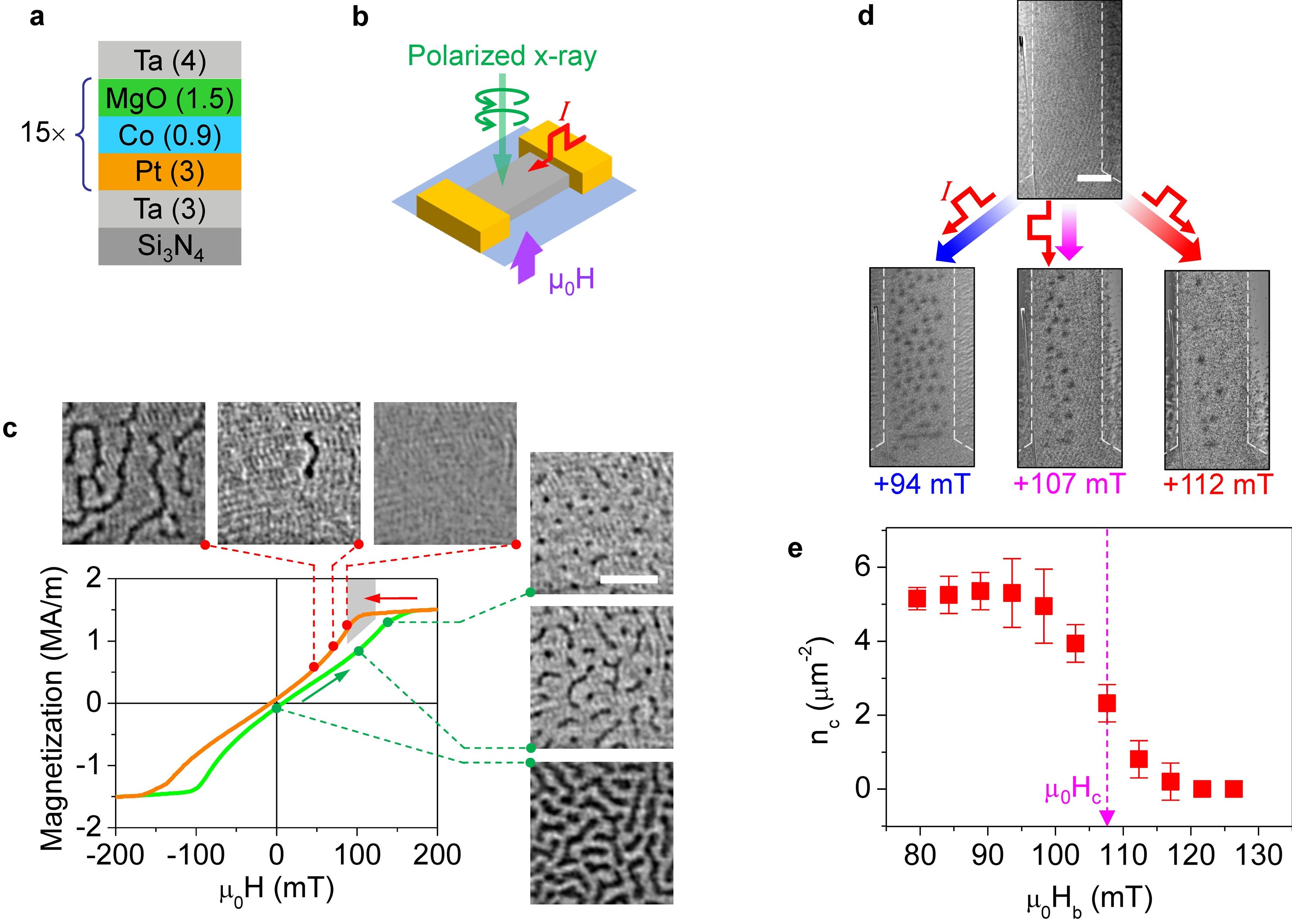}
\caption{\textbf{Experimental setup and current-driven skyrmion writing experiments.}
\textbf{(a)} Schematic of the multilayer thin film used in this work
(layer thickness in nm in parentheses). \textbf{(b)} Schematic of
the MTXM experimental setup. \textbf{(c)} Hysteresis loop for OP magnetization,
$M\left(\mu_{0}H\right)$, with corresponding MTXM images (scale bar:
1 $\mu$m) at OP fields indicated by vertical dashed lines. Top three
images (right to left) show the formation of chiral magnetic textures
as $\mu_{0}H$ is reduced from positive saturation ($+\mu_{0}H_{{\rm s}}$),
while the bottom three (bottom to top) show their shrinking and disappearance
as $\mu_{0}H$ is increased from zero to $+\mu_{0}H_{{\rm s}}$. The
shaded grey field region, corresponding to a uniformly magnetized
state, indicates the base field range for experiments in (d-e) and
\ref{Fig:3_SkDel-Expts}. \textbf{(d)} Typical examples of MTXM images
(scale bar: 1~$\mu$m) of the wire before and after the current pulse.
Top image shows the uniformly magnetized initial state at $\mu_{0}H_{{\rm b}}$.
Bottom images show the final states (i.e., after a current pulse)
at each $\mu_{0}H_{{\rm b}}$. \textbf{(e)} Plot of the density, $n_{{\rm c}}$,
of skyrmions nucleated as per (d) against the base field $\mu_{0}H_{{\rm b}}$.
The inferred crossover field (see \ref{Fig:3_SkDel-Expts} and associated
text) is labelled as $\mu_{0}H_{{\rm c}}$.}
\label{Fig:1_SkWrite-Expts}
\end{figure*}

This work was performed on multilayer stacks of {[}Pt(3)/Co(0.9)/MgO(1.5){]}$_{15}$
(\ref{Fig:1_SkWrite-Expts}a, hereafter Pt/Co/MgO), which were sputtered
on X-ray transparent Si$_{3}$N$_{4}$ membranes, and fabricated into
2~$\mu$m wide wires (see Methods). The asymmetric Pt/Co/MgO stack
hosts a sizeable interfacial Dzyaloshinskii-Moriya interaction (DMI),
which can stabilize Néel-textured skyrmions at room temperature (RT)
\citep{Boulle.2016,Juge.2019}. To observe sub-100~nm magnetic skyrmions
in Pt/Co/MgO multilayers, we employed full field magnetic transmission
soft x-ray microscopy (MTXM) at the Advanced Light Source (XM-1, BL6.1.2)
\citep{Chao.2005}. Out-of-plane (OP) geometry (\ref{Fig:1_SkWrite-Expts}b)
is employed to detect the OP magnetization through X-ray magnetic
circular dichroism. The imaging results, acquired at RT with in situ
OP magnetic field $\mu_{0}H$, are complemented by micromagnetic simulations
using magnetic parameters determined from magnetometry measurements
(see Methods).

We begin by examining the equilibrium magnetization $M(\mu_{0}H)$
of the Pt/Co/MgO multilayer, and corresponding MTXM images obtained
over a wide wire pad (\ref{Fig:1_SkWrite-Expts}c). The sheared hysteresis
is characteristic of a multi-domain configuration at remanence as
confirmed by the labyrinthine zero field (ZF) MTXM image. As $\mu_{0}H$
is swept down from positive saturation (orange branch in \ref{Fig:1_SkWrite-Expts}c),
stripe domains nucleate at $\sim70$~mT, which propagate to form
a labyrinthine state. Meanwhile, on the upsweep (green branch in \ref{Fig:1_SkWrite-Expts}c),
increasing $\mu_{0}H$ beyond zero results in the labyrinthine state
transforming to oppositely magnetized stripes, which eventually shrink
to a sparse array of sub-100~nm skyrmions. For the subsequent electrical
writing experiments, the used magnetic fields correspond to the shaded
region of \ref{Fig:1_SkWrite-Expts}c, wherein the initial MTXM images
are uniformly magnetized. Slight discrepancies between device level
imaging and film-level $M(\mu_{0}H)$ results may be attributed to
the difference in the numbers of nucleation sites between the patterned
wire\citep{Woo.2017} and the film \citep{Soumyanarayanan.2017}.

\section*{Electrical Writing of Skyrmions\label{sec:SkWriting}}

\begin{figure*}
\includegraphics[width=0.85\textwidth]{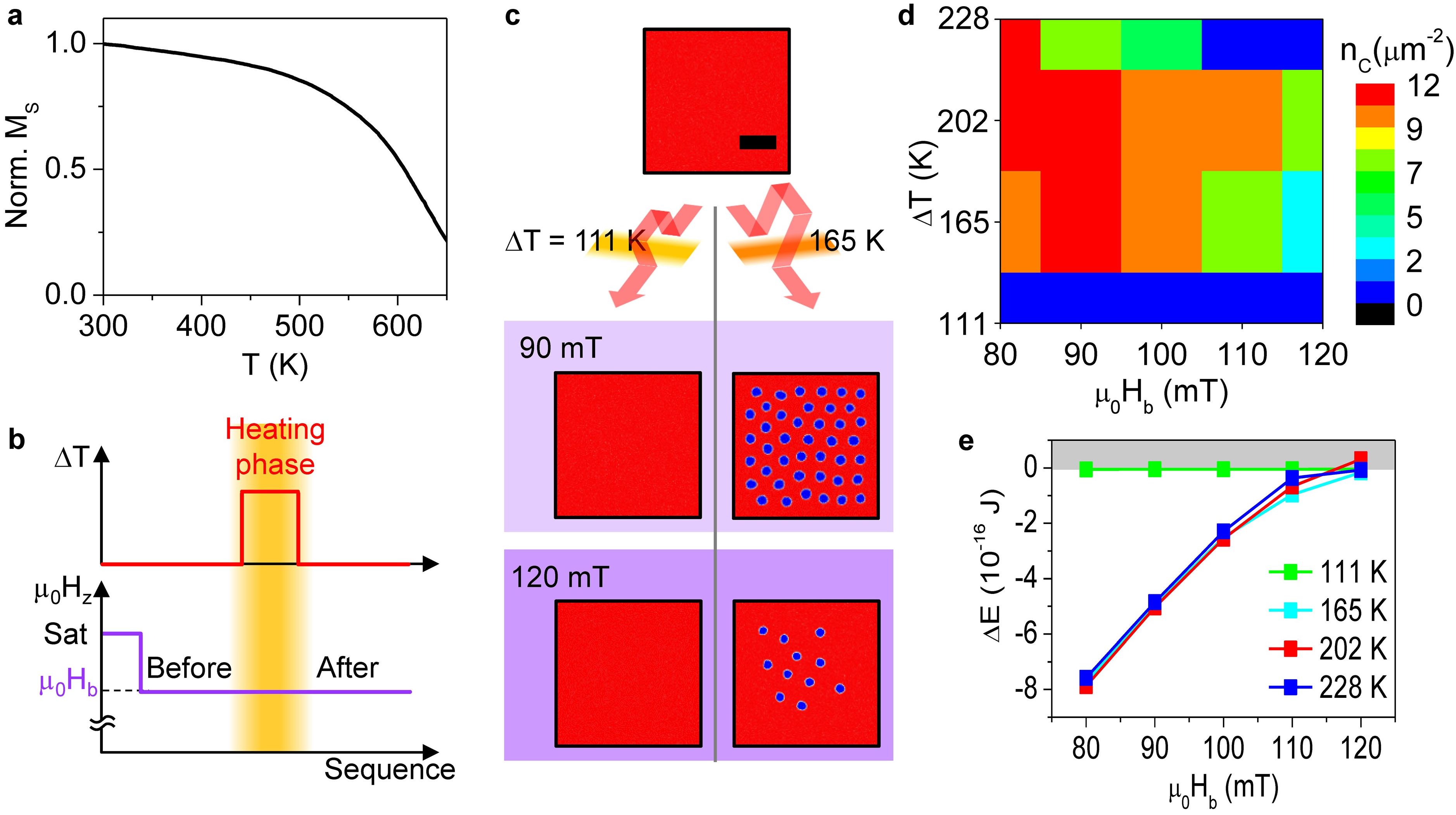}
\caption{\textbf{Simulations of heat-induced skyrmion nucleation.} \textbf{(a)}
Measured variation of saturation magnetization ($M_{{\rm S}}$) with
temperature after normalizing to its 300~K value. This data is used
along with scaling relations of magnetic parameters as inputs for
Joule heating simulations. \textbf{(b)} Schematic recipe used for
micromagnetic simulations of Joule heating effects. \textbf{(c)} Typical
examples of simulation results for \emph{$\Delta T$} = 111~K, 165~K
and $\mu_{0}H_{{\rm b}}$ = 90~mT, 120~mT, respectively (scale bar:
0.5~$\mu$m). Simulations are initialized at $\mu_{0}H>\mu_{0}H_{{\rm s}}$,
and the uniform state is brought to $\mu_{0}H_{{\rm b}}$ (c, top).
During the heating phase (b, centre), the simulated temperature is
raised by $\Delta T$ and magnetic parameters are rescaled (see text,
\ref{Fig:4_SkDel-Sims}a). Skyrmions are nucleated in some cases (c,
right), and persist when RT conditions are restored. \textbf{(d)}
Color plot showing the nucleated skyrmion density, $n_{{\rm c}}$,
as a function of $\Delta T$ and $\mu_{0}H_{{\rm b}}$ (the latter
c.f. \ref{Fig:1_SkWrite-Expts}e). \textbf{(e)} Micromagnetic energy
difference, $\Delta E$, between final (skyrmion nucleated) and initial
(uniform) states for various \emph{$\Delta T$} and $\mu_{0}H_{{\rm b}}$.
The final state is more stable if $\Delta E<0$ (note: \emph{$\Delta E=0$}
for $\Delta T=111$~K as the initial and final states are identical).}
\label{Fig:2_SkWrite-Sims}
\end{figure*}

For electrical writing experiments, we begin by saturating the sample
at large positive field, and lower it to a base field $\mu_{0}H_{{\rm b}}$
ranging from 80 to 126~mT (\ref{Fig:1_SkWrite-Expts}c, shaded grey
region) wherein the sample retains uniform magnetization (\ref{Fig:1_SkWrite-Expts}d:
top). Under these conditions, the injection of a single current pulse
(width $\sim30$~ns, amplitude $\sim6.8\times10^{11}$~A/m$^{2}$)
results in the nucleation of skyrmions in the wire device (\ref{Fig:1_SkWrite-Expts}d:
bottom). We observe that the number of nucleated skyrmions varies
considerably with $\mu_{0}H_{{\rm b}}$ as seen across the MTXM images
(\ref{Fig:1_SkWrite-Expts}d: bottom). The skyrmion nucleation is
quantified in \ref{Fig:1_SkWrite-Expts}e by measuring the density
of created skyrmions, $n_{{\rm c}}$, with respect to $\mu_{0}H_{{\rm b}}$.
Below $\mu_{0}H_{{\rm b}}\sim100$~mT, the current pulse nucleates
a dense skyrmion array, with density $n_{{\rm c}}>$ 5~$\mu$m$^{-2}$
(e.g.. left in \ref{Fig:1_SkWrite-Expts}d). Above $\mu_{0}H_{{\rm b}}\sim$100~mT,
$n_{{\rm c}}$ drops sharply with increasing $\mu_{0}H_{{\rm b}}$
(middle in \ref{Fig:1_SkWrite-Expts}d), and reaches zero at \textasciitilde 120~mT.
Notably, the modulation of the electrical nucleation of skyrmions
by the external magnetic field is distinct from those reported in
previous works \citep{Jiang.2015,Finizio.2019,Buttner.2017,Woo.2018,Lemesh.2018}.

\begin{figure*}
\includegraphics[width=0.6\textwidth]{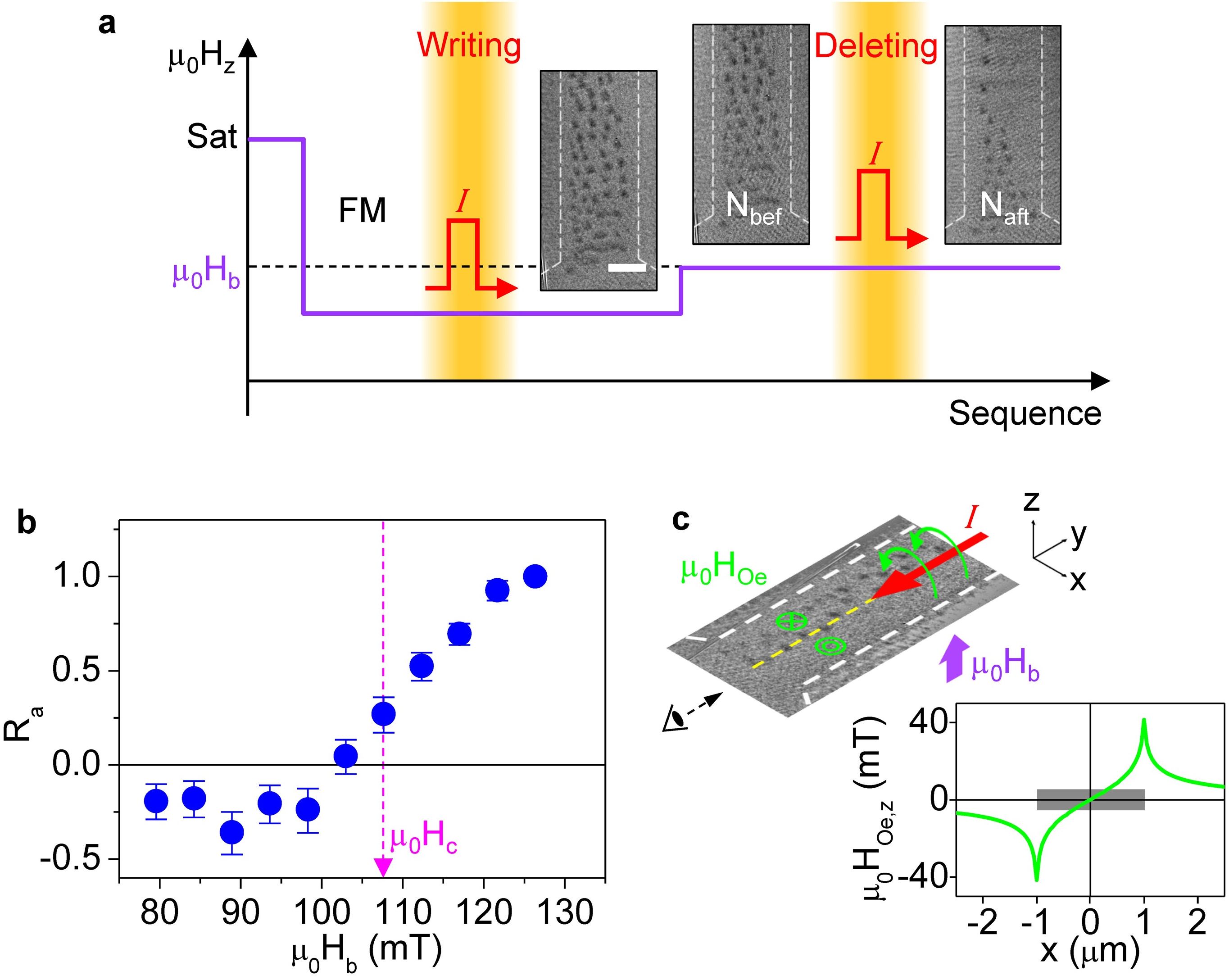}
\caption{\textbf{Current-driven skyrmion deletion experiments.} \textbf{(a)}
Recipe used for skyrmion deletion, with representative MTXM images
(scale bar: 1 \textmu m) as insets. \textbf{(b)} Plot of the skyrmion
annihilation rate, $R_{{\rm a}}\left(\equiv1-N_{\text{aft}}/N_{\text{bef}}\right)$,
with varying base field $\mu_{0}H_{{\rm b}}$. The inferred crossover
field, $\mu_{0}H_{{\rm c}}$, is indicated (c.f. \ref{Fig:1_SkWrite-Expts}e,
see caption). \textbf{(c)} Plot of the expected Oersted field profile
(inset shows schematic) generated by a current applied along $-\widehat{y}$
(red arrow) as a function of the transverse position $x$ across the
wire.}
\label{Fig:3_SkDel-Expts}
\end{figure*}

Lateral currents within an asymmetric stack (e.g., Pt/Co/MgO) may
produce a large SOT \citep{Miron.2011}. Such SOT governs skyrmion
motion and, in some recent works, induces skyrmion nucleation \citep{Finizio.2019,Buttner.2017,Woo.2018}.
However, the SOT-driven writing mechanisms cannot explain our results
for the following reasons. First, our device does not have any geometric
constrictions, positioned defects, or in-plane magnetic fields required
for the SOT writing of skyrmions \citep{Finizio.2019,Buttner.2017}.
Second, unlike the peculiar bipolar pulse waveform used to achieve
the dynamic SOT nucleation \citep{Woo.2018}, we use a conventional
single pulse with a rectangular profile. Crucially, as we will show
in \ref{Fig:3_SkDel-Expts}, the same pulse can be used for deletion
of skyrmions which clearly rules out vectorial mechanisms associated
with SOT. On the other hand, a plausible explanation may be the Joule
heating effect and associated thermodynamics, which were previously
shown to drive transitions from stripe domains to skyrmions \citep{Lemesh.2018}.
For our case, simulation and analytical modelling suggest that current
pulses used may induce a temperature rise $\Delta T\sim$ 220 K\citep{Fangohr.2011}.

To investigate the plausibility of heat-induced skyrmion nucleation,
we performed micromagnetic simulations (see Methods). Typically, such
simulations account for temperature effects solely by introducing
randomly fluctuating fields \citep{Vansteenkiste.2014}. However the
$\Delta T$ would additionally change the magnetic parameters. To
consider this effect, we measured the saturation magnetization $M_{{\rm S}}(T)$
over 300-650~K (\ref{Fig:2_SkWrite-Sims}a), and estimated the micromagnetic
parameters at elevated temperatures using scaling relations with magnetization
\citep{Nembach.2015,Moreno.2016,Tomasello.2018}: exchange stiffness
$A\propto M_{{\rm S}}^{1.8}$, interfacial DMI $D\propto M_{{\rm S}}^{1.8}$
and uniaxial anisotropy $K_{{\rm u}}\propto M_{{\rm S}}^{3.03}$.
The micromagnetic simulations were then performed following the recipe
depicted in \ref{Fig:2_SkWrite-Sims}b (details in Supporting Information
(SI)). In particular, SOT effects were not included. The system is
first initialized to a uniformly magnetized state at $\mu_{0}H_{{\rm b}}$
(\ref{Fig:2_SkWrite-Sims}c: top), subjected to a heating phase with
rescaled parameters, and then restored to RT conditions (\ref{Fig:2_SkWrite-Sims}c:
bottom).

\ref{Fig:2_SkWrite-Sims}c shows the simulation results for representative
values of $\Delta T$ (111, 165~K) and $\mu_{0}H_{{\rm b}}$ (90,
120~mT). For $\Delta T$ = 165~K (\ref{Fig:2_SkWrite-Sims}c, right),
skyrmions are nucleated by the heat pulse, and their number decreases
as $\mu_{0}H_{{\rm b}}$ increases. In contrast, the smaller heat
pulse $\Delta T$ = 111 K (\ref{Fig:2_SkWrite-Sims}c, left) does
not create any skyrmions regardless of $\mu_{0}H_{{\rm b}}$. \ref{Fig:2_SkWrite-Sims}d
shows in further detail the simulated $n_{c}$ with $\Delta T$ (100-250~K)
and $\mu_{0}H_{{\rm b}}$ (80-120~mT). We find that a threshold $\Delta T>$
150~K is required to nucleate any skyrmions, corresponding to a scaled
magnetization $M_{s}(300+\Delta T)\sim0.9M_{s}(300\,{\rm K})$. Above
this threshold, the field dependence of $n_{{\rm c}}$ is qualitatively
consistent with the experimental data (\ref{Fig:1_SkWrite-Expts}e).
Overall, this indicates that Joule heating induced by current pulses
is a viable explanation of the observed nucleation of skyrmions and
the dependence of $n_{{\rm c}}$ on $\mu_{0}H_{{\rm b}}$.

Micromagnetic simulations further help to elucidate several aspects
of heat-induced skyrmion energetics. First, \ref{Fig:2_SkWrite-Sims}e
shows the energy difference, $\Delta E$, between the final and initial
(ambient) states with varying $\Delta T$ and $\mu_{0}H_{{\rm b}}$.
Notably $\Delta E$ is negative for most cases, with negligible variation
over $\Delta T$ of 165-228~K, i.e., the skyrmion formation reduces
the total energy with respect to that of the uniform state. This indicates
that skyrmion nucleation is thermodynamically favoured (explored further
in Conclusions), except near saturation ($\mu_{0}H_{{\rm b}}=$ 120~mT)
wherein skyrmions may be metastable. Next, we note a slight qualitative
difference in the nucleation phenomenology with $\Delta T$ (details
in SI). For intermediate $\Delta T$ ($<220$~K), skyrmions are formed
during the heating phase. However, for higher $\Delta T$ ($>220$~K),
i.e., $M_{s}(T)<0.8M_{s}(300\,{\rm K})$, fluctuations during the
heat phase result in random magnetization, and skyrmions are formed
only when the heat is turned off. Finally, simulations allow for a
sequential study of the two heat-induced effects: thermal fluctuations
and rescaling of magnetic parameters. Crucially, we find that no skyrmions
are nucleated at any $\mu_{0}H_{{\rm b}}$ or $\Delta T$ if magnetic
parameters are not rescaled (see SI). Together, these simulation insights
strongly point to the thermodynamic origin of the observed nucleation
of skyrmions. Finally, while the simulation results in \ref{Fig:2_SkWrite-Sims}
are grain-free, potential effects of pinning sites on heating and
skyrmion nucleation are discussed in the SI. 

\section*{Electrical Deletion of Magnetic Skyrmions}

\begin{figure*}
\includegraphics[width=0.9\textwidth]{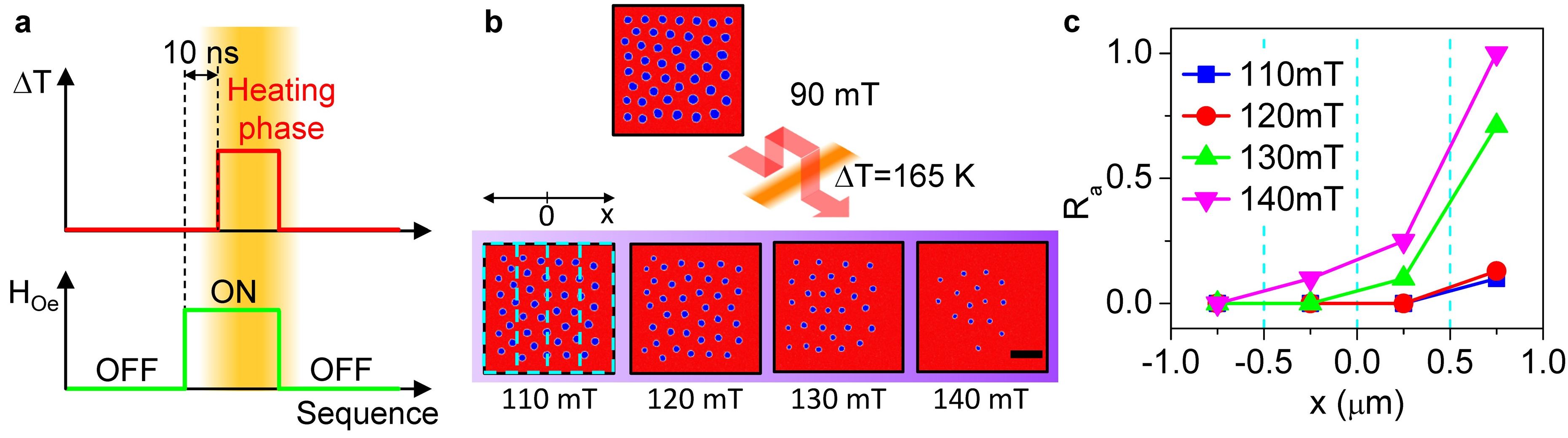}
\caption{\textbf{Simulations of Oersted field-induced skyrmion annihilation.}
\textbf{(a)} Schematic recipe used for simulating Oersted field effect
with Joule heating (details in SI). The current pulse is modelled
as inducing an Oersted field (profile in \ref{Fig:3_SkDel-Expts}c)
for the entire 30~ns duration (bottom) and Joule heating for the
latter 20~ns (top). \textbf{(b)} Simulated magnetization before (top)
and after (bottom) the base field is raised to $\mu_{0}H_{{\rm b}}$
(110-140~mT), and the recipe in (a) is used. Skyrmions are annihilated
in some cases (bottom right). \textbf{(c)} Skyrmion annihilation rate
$R_{{\rm a}}$ plotted as a function of transverse position, $x$,
across the wire for several base fields $\mu_{0}H_{{\rm b}}$. $R_{{\rm a}}$
is determined by the number of skyrmions in the dashed boxes in (b).
The four sections in (c) divided by the cyan dashed lines correspond
to the dashed boxes in (b).}
\label{Fig:4_SkDel-Sims}
\end{figure*}

Meanwhile, we find that a slight modification of the skyrmion nucleation
recipe can be used to annihilate skyrmions, as schematically shown
in \ref{Fig:3_SkDel-Expts}a along with representative MTXM images.
Here, the initial state consists of skyrmions created using the writing
recipe at $\mu_{0}H$ = 94~mT (\ref{Fig:3_SkDel-Expts}a: left).
The magnetic field is then changed to a specific base field $\mu_{0}H_{{\rm b}}$
(\ref{Fig:3_SkDel-Expts}a: centre). The second pulse (deleting) is
then injected to the wire, and the numbers of skyrmions before ($N_{\text{bef}}$)
and after ($N_{\text{aft}}$) the current pulse are counted. Here
we emphasize that identical current pulses are used for writing and
deletion. As shown in \ref{Fig:3_SkDel-Expts}a, the number of skyrmions
is reduced by $\sim50\%$ following the applied current pulse. Note
that the skyrmion annihilation has distint spatial asymmetry: skyrmion
annihilation is more prominently on the right side of the wire. We
quantify the efficacy of deletion by the annihilation rate $R_{{\rm a}}\left(\equiv1-N_{\text{aft}}/N_{\text{bef}}\right)$.
\ref{Fig:3_SkDel-Expts}b shows a plot of $R_{{\rm a}}$ as a function
of $\mu_{0}H_{{\rm b}}$. For lower $\mu_{0}H_{{\rm b}}$ ($<100$~mT),
$R_{{\rm a}}$ maintains a nearly constant negative value, indicating
that skyrmions are instead nucleated by the current pulse (see \ref{Fig:1_SkWrite-Expts}e).
Above $\sim100$~mT, $R_{{\rm a}}$ increases rapidly, consistent
with the annihilation of existing skyrmions, and reaches unity near
saturation ($\sim130$~mT). The observation that the same current
pulse drives nucleation at lower $\mu_{0}H_{{\rm b}}$ and annihilation
at higher $\mu_{0}H_{{\rm b}}$ suggests the existence of a cross-point
field $\mu_{0}H_{{\rm c}}$ ($\sim107$~mT), corresponding to a dynamic
equilibrium between these two phenomena.

The striking spatial asymmetry of skyrmion deletion (\ref{Fig:3_SkDel-Expts}a:
right), mainly affecting the right side of the wire, is suggestive
of an Oersted field driven phenomenon. Notably, reversing the current
polarity flips the spatial asymmetry of skyrmion annihilation (SI),
further supporting the Oersted field effect. \ref{Fig:3_SkDel-Expts}c
shows the Oersted field profile during the current pulse, calculated
by superposing the fields generated by current segments through the
wire cross-section (SI). The Oersted field is sizable (up to $\pm40$~mT)
compared to $\mu_{0}H_{{\rm b}}$ (80-130~mT), and introduces considerable
spatial asymmetry in the skyrmion distribution by increasing (decreasing)
the net perpendicular field on the right (left) of the wire as shown
in \ref{Fig:1_SkWrite-Expts}e and \ref{Fig:3_SkDel-Expts}b. Near
the cross-point field $\mu_{0}H_{{\rm c}}\sim107$~mT, both nucleation
and annihilation are likely equally effective, resulting in the clear
spatial segregation of nucleation and annihilation as observed in
\ref{Fig:1_SkWrite-Expts}d and \ref{Fig:3_SkDel-Expts}a. Therefore,
our scheme has distinctive merits when operating near $\mu_{0}H_{{\rm c}}$
as one can selectively switch between nucleation and annihilation
at a given spatial location simply by reversing the current polarity.

To elucidate the annihilation process, micromagnetic simulations were
performed following the recipe depicted in \ref{Fig:4_SkDel-Sims}a
(details in SI). Notably, while a current generates an instantaneous
Oersted field, generating appreciable $\Delta T$ (c.f. \ref{Fig:2_SkWrite-Sims})
requires a few tens of nanoseconds \citep{Kim.2008}. Our simulation
recipe accounts for this lag by using an Oersted field (profile shown
in \ref{Fig:3_SkDel-Expts}c) for the entire 30~ns pulse duration,
followed by the heating phase introduced after a 10~ns delay. \ref{Fig:4_SkDel-Sims}b
shows simulation results at several base fields $\mu_{0}H_{{\rm b}}$
of 110-140~mT, while \ref{Fig:4_SkDel-Sims}c displays the resulting
annihilation rate $R_{{\rm a}}$ with respect to the lateral position
$x$ (defined in \ref{Fig:4_SkDel-Sims}b). Consistent with experiments
(\ref{Fig:3_SkDel-Expts}b), the simulated $R_{{\rm a}}$ with $\mu_{0}H_{{\rm b}}$
and the deletion is more prominent on the right side of the wire.
Meanwhile, we note that identical current pulses used in the skyrmion
writing experiments would produce similar Oersted fields. Therefore,
we have verified that such an Oersted field does not affect the simulated
nucleation results (see SI). The contrasting influence of the Oersted
field in the two cases is likely due to differences in initial conditions
and the base field magnitude.

Interestingly, the simulations allow us to analyse the annihilation
process in stepwise fashion (c.f. \ref{Fig:4_SkDel-Sims}a) to disentangle
the various effects at play. The crucial finding is that all annihilation
events in \ref{Fig:4_SkDel-Sims} occur within the first 10~ns, wherein
$\Delta T$ is assumed to be nearly zero (SI). No additional annihilation
is observed at elevated temperatures with or without the Oersted field.
In conjunction with the negative $\mathrm{\Delta}E$ in \ref{Fig:2_SkWrite-Sims}e,
this suggests that Oersted field and Joule heating effects have counteracting
outcomes in (de)stabilizing skyrmions. 

\begin{figure*}
\includegraphics[width=0.65\textwidth]{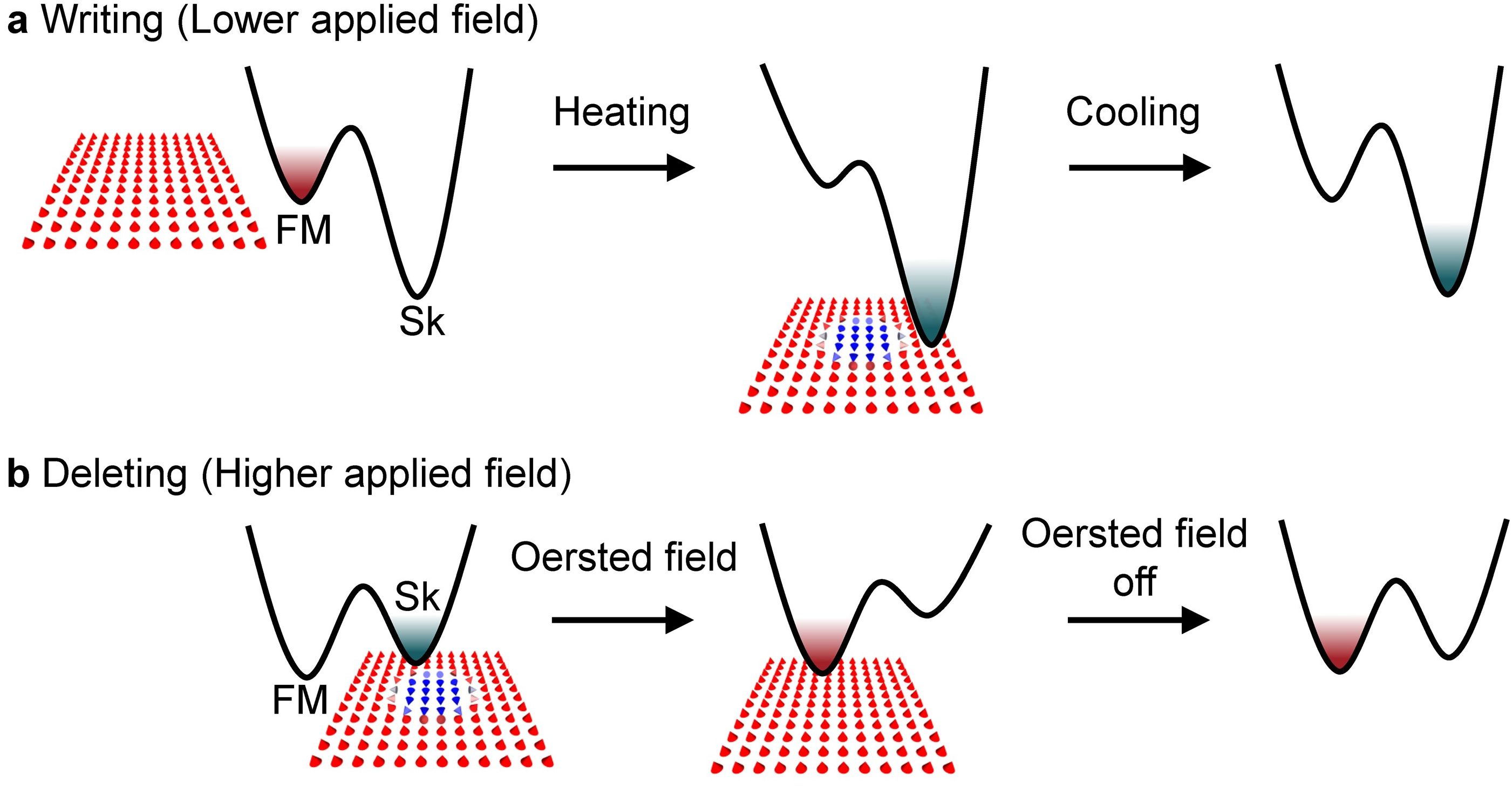}
\caption{\textbf{Schematic energetics for writing and deleting mechanism.}
\textbf{(a)} Skyrmions are written onto the uniform state by injecting
a current pulse at lower $\mu_{0}H_{{\rm b}}$, wherein they are energetically
favored. Joule heating induced by the current pulse drives their nucleation
by lowering the energy barrier. \textbf{(b)} Deletion is effective
at higher $\mu_{0}H_{{\rm b}}$ where the uniform and skyrmion states
are comparable in energy. An Oersted field added into the base field
raises the energy of the skyrmion state, and thereby triggers annihilation
of the skyrmion by injecting a current pulse.}
\label{Fig:5_WriteDel-Energetics}
\end{figure*}

Notably, while thermal fluctuations have also been suggested to induce
annihilation of skyrmions \citep{Legrand.2017}, for our case, Joule
heating induces skyrmion nucleation and prevents annihilation through
rescaling of magnetic parameters over a wide temperature range. Therefore,
the time delay between the Oersted field and heating effects, introduced
to reflect experimental reality, is crucial to enabling skyrmion deletion.
Overall, while the inclusion of a more realistic current profile and
material granularity may enable better quantitative agreement with
experiments, the simulation results shown here establish the key mechanistic
insights needed to interpret the skyrmion creation and annihilation
observed in our wire devices.

\section*{Summary and Outlook}

\ref{Fig:5_WriteDel-Energetics}a,b summarizes the energetics of skyrmion
nucleation and annihilation via Joule heating and Oersted field effects.
First, at lower base fields, the skyrmion state becomes favorable
relative to the uniform state, but nucleation is prevented by an energy
barrier (\ref{Fig:5_WriteDel-Energetics}a: left). Magnetic parameters
at elevated temperatures resulting from Joule heating reduce this
barrier, and, incidentally, further lower the skyrmion energy, thereby
enabling skyrmion nucleation (\ref{Fig:5_WriteDel-Energetics}a: center,
right). The number of skyrmions formed depends on the energy difference
of the uniform and skyrmion states, which may be controlled by adjusting
the base field. Meanwhile, at higher base fields, the uniform state
is comparable in energy with the skyrmion (\ref{Fig:5_WriteDel-Energetics}b:
left), and the energy barrier now prevents skyrmion annihilation.
However, the addition of an Oersted field pulls up the skyrmion energy,
i.e., further destabilizes skyrmions, resulting in their annihilation
regions with increased effective field

In summary, we have described our observations of the writing and
deleting of skyrmions by lateral currents applied to a two-terminal
wire device. A lateral electric current induces two thermodynamic
effects: Joule heating and magnetic field. Our work harnesses the
distinct spatial and temporal characteristics of these effects to
write and delete skyrmions with efficacies that may be tuned by the
external bias field. On one hand, Joule heating drives the nucleation
of skyrmions, inherently stable at lower fields, by modifying magnetic
parameters at elevated temperatures. On the other hand, the Oersted
field preceding Joule heating enables annihilation of skyrmions in
regions of higher fields. Together, our experiments and simulations
provide a detailed energetic picture of these mechanisms that are
generalizable to other skyrmion-hosting materials.

Our writing and deleting schemes are uniquely promising from a scalability
perspective, considering the simplicity of device design and electrical
inputs needed to implement the schemes. Their immense practical value
will inspire immediate efforts towards deterministic and field-free
skyrmion manipulation in such devices \citep{Finizio.2019,Buttner.2017,Woo.2018}.
First, while both Joule heating and Oersted field effects may coexist
and counteract, one could ensure for either of these to dominate by
varying the length and amplitude of the current pulse \citep{Finizio.2019}.
Next, the requisite of external fields could be realized by appropriate
use of magnetostatic effects \citep{Zeissler.2017,Juge.2018,Ho.2019,Finocchio.2016}.
While these schemes could be used together in skyrmionic racetrack
devices \citep{Fert.2017,Finizio.2019}, each scheme can also be utilized
for independently manipulating skyrmions. For example, the spatially
selective nature of Oersted field-induced annihilation could be used
to realize fabrication-free logic operations \citep{Kang.2016}. Finally,
Joule heating presents itself as a promising technique to emulate
a skyrmion bath or reservoir, providing a timely experimental platform
for recently proposed novel computing architectures \citep{Pinna.2018,Prychynenko.2018}.

\section*{Methods}

\textbf{\small{}Micromagnetic Simulations.}{\small{} Simulations were
performed using mumax$^{3}$ software package \citep{Vansteenkiste.2014}.
The effective medium theory was used to examine an equivalent reduced
stack while retaining interlayer interactions \citep{Woo.2016}. The
magnetic parameters were as follows (effective medium parameters in
parenthesis): $A=2.4\times10^{-11}$~J/m ($4.0\times10^{-12}$~J/m),
$M_{{\rm S}}=1.59\times10^{6}$~A/m ($2.64\times10^{5}$~A/m), $K_{{\rm u}}=1.63\times10^{6}$~J/m$^{3}$
($K_{{\rm u}}=5.21\times10^{4}$~J/m$^{3}$) and $D=1.76\times10^{-3}$~J/m$^{2}$
($2.93\times10^{-4}$~J/m$^{2}$). These parameters were determined
using procedures in previous works \citep{Soumyanarayanan.2017,Moreau-Luchaire.2016,Woo.2016},
and are in line with published results on similar stacks \citep{Boulle.2016}.}{\small\par}

\textbf{\small{}Joule Heating.}{\small{} Experimental determination
of $\Delta T$ of the wire is challenging due to the low thermal conductivity
of the membrane (substrate) and the short duration of the current
pulse (30~ns). Therefore, $\Delta T$ is estimated using analytical
model\citep{Fangohr.2011} with an assumption all charge is flowing
the Pt layers, and is found to be $\sim220$~K. Correspondingly,
micromagnetic simulations were performed at elevated temperatures
(up to $\sim600$~K) to emulate the effect of heating. The elevated
temperatures were reflected in: (a) the fluctuating thermal field
in mumax$^{3}$, and (b) rescaling of magnetic parameters to reflect
their values at the elevated temperatures.}{\small\par}

\textbf{\small{}Film Deposition and Device Fabrication.}{\small{}
Multilayer stacks of Ta(4)/ {[}Pt(3)/Co(0.9)/MgO(1.5){]}$_{15}$/Ta(4)/Ru(5)
(nominal layer thicknesses in nm in parentheses) were deposited on
200~mm Si/SiO$_{2}$ (20,000) wafers by ultrahigh-vacuum magnetron
sputtering at RT using a Timaris™ UHV system (base pressure $2\times10^{-8}$~Torr)
manufactured by Singulus Technologies AG. The active stack was repeated
15 times for optimal XMCD contrast, and the film was simultaneously
deposited on 200~nm thick X-ray transparent Si$_{3}$N$_{4}$ membranes
for the MTXM measurement. Nanowire devices for the MTXM measurement
were fabricated on the Si$_{3}$N$_{4}$ membranes using electron
beam lithography (Elionix™ tool), ion-beam etching (Oxford CAIBE™),
and Ta(5)/Au(100)/Ru(20) electrode deposition (Bestec Chiron™). Completeness
of resist removal was ensured using an Axcelis™ O$_{2}$ stripping
tool. }{\small\par}

\textbf{\small{}Magnetization measurements}{\small{} were performed
over 300-700~K using the EZ11 vibrating sample magnetometer (VSM)
from MicroSense™ using the stacks deposited on the Si/SiO$_{2}$ substrates. }{\small\par}

\textbf{\small{}MTXM Measurements.}{\small{} A full-field magnetic
transmission soft x-ray microscope (MTXM; XM-1, BL6.1.2) located at
the Advanced Light Source was used to image sub-100~nm magnetic skyrmions
in the nanowires. Imaging was performed in out-of-plane geometry at
the Co $L_{3}$ X-ray absorption edge ($\sim778$~eV). A nanosecond
pulse generator (Agilent 81150A) was used to apply current pulses
to the nanowires.}{\small\par}

\begin{center}
{\small{}\rule[0.5ex]{0.6\columnwidth}{0.5pt}}{\small\par}
\par\end{center}

\smallskip{}
\textbf{Acknowledgements. }Works at the ALS were supported by U.S.
Department of Energy (DE-AC02-05CH11231). S.-G.J. was supported by
the National Research Foundation of Korea (NRF, Korea) grant funded
by the Korea government (MSIT) (2020R1C1C1006194). M.-Y.I. acknowledges
support by Lawrence Berkeley National Laboratory through the Laboratory
Directed Research and Development (LDRD) Program. This work in Singapore
was supported by the SpOT-LITE programme (Grant Nos. A1818g0042, A18A6b0057),
funded by Singapore's RIE2020 initiatives, and by the Pharos Skyrmion
programme (Grant No. 1527400026) funded by A{*}STAR, Singapore. We
also acknowledge the support of the National Supercomputing Centre
(NSCC), Singapore for computational resources. K.-S.L. and D.-H.J.
were supported by NRF of Korea grant funded MSIT (NRF-2019R1A2C2002996,
NRF-2016M3D1A1027831, and NRF-2019K1A3A7A09033400). J.-I.H. was supported
by NRF of Korea grant funded MSIT (2020R1A2C2005932).

\phantomsection
\addcontentsline{toc}{section}{\refname} \bibliography{JouleHtg,SkFORC}

\clearpage\normalsize 
\clearpage 
\onecolumngrid 

\begin{center} 
\textbf{\Large Supplementary Information} 
\end{center} 

\setcounter{section}{0}
\setcounter{figure}{0}
\setcounter{table}{0} 
\setcounter{equation}{0}

\setcounter{secnumdepth}{4}
\renewcommand\thesection{S\arabic{section}} 
\renewcommand{\theparagraph}{S\arabic{section}\alph{paragraph}} \makeatletter\@addtoreset{paragraph}{section}\makeatother 
\makeatletter\def\p@paragraph{}\makeatother 
\renewcommand{\thefigure}{S\arabic{figure}}
\renewcommand{\theequation}{S\arabic{equation}} \renewcommand{\thetable}{S\arabic{table}}

\renewcommand{\ref}[1]{\autoref{#1}} 
\renewcommand{\figureautorefname}{Fig.} 
\renewcommand{\equationautorefname}{Eqn.} 
\renewcommand{\tableautorefname}{Tbl.} 
\renewcommand{\sectionautorefname}{\S} 
\renewcommand{\subsectionautorefname}{\S} 
\renewcommand{\paragraphautorefname}{\S}

\setcounter{tocdepth}{1} 
\makeatletter\def\l@section{\@dottedtocline{1}{0.6em}{1.5em}}\makeatother \makeatletter\def\l@paragraph{\@dottedtocline{4}{1.5em}{1.8em}}\makeatother \makeatletter\def\l@figure{\@dottedtocline{1}{0.6em}{1.8em}}\makeatother

\linespread{1.25} 
\def\arraystretch{1.5} 
\setlength{\parskip}{1.5ex plus0.2ex minus0.2ex} 
\setlength{\abovecaptionskip}{4pt}
\setlength{\belowcaptionskip}{-4pt} 
\setlength{\abovedisplayskip}{0ex}
\setlength{\belowdisplayskip}{0ex} 
\setlength{\abovedisplayshortskip}{0ex}
\setlength{\belowdisplayshortskip}{0ex}

\titleformat{\section}{\large\bfseries\scshape\filcenter}{\thesection.}{1em}{#1}[{\titlerule[0.5pt]}] 
\titlespacing*{\section}{0pt}{1ex}{1ex} 
\titleformat{\subsection}{\bfseries\sffamily}{\thessubsection.}{0em}{#1} 
\titlespacing{\subsection}{0pt}{0.5ex}{0.5ex} 
\titleformat{\paragraph}[runin]{\sffamily\bfseries}{}{-1.2em}{#1.} 
\titlespacing*{\paragraph}{1.25em}{2ex}{0.4em}[]

\section{Simulating Joule Heating Effects\label{sec:Sims_JHtg}}

\begin{figure}[h]
\centering \includegraphics[width=0.6\textwidth]{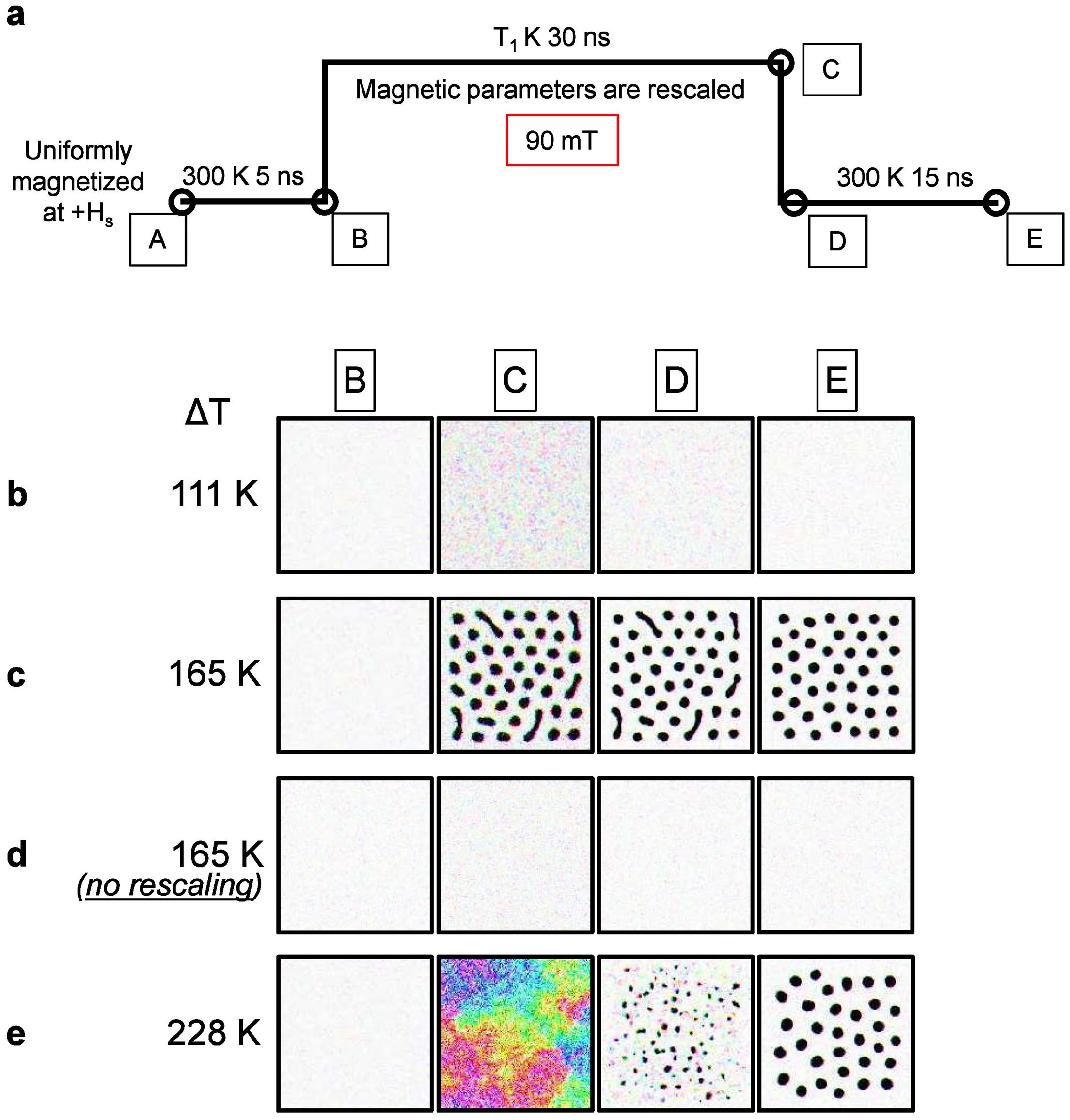}
\caption{\textbf{Simulations of Joule heating-driven skyrmion writing.} \textbf{(a)}
Schematic of the simulation recipe used at a base field of 90~mT
to emulate a temperature rise $\Delta T$ for the current pulse duration,
with rescaling of magnetic parameters (see main text) during the pulse.
Labels A to E identify various instances within the simulation recipe.
\textbf{(b}-\textbf{e)} Grayscale snapshots of simulated out-of-plane
magnetization at instances B to E {[}defined in (a){]} for (b) $\Delta T$
= 111~K, (c) $\Delta T$ = 165~K, (d) $\Delta T$ = 165~K (without
rescaling of magnetic parameters) and (e) $\Delta T$ = 228~K. While
B is identical for all cases, the intermediate (C, D) and final (E)
configurations vary, and show skyrmion formation in some cases.}
\label{Fig:S1_Sims-JHtg-SkWrite}
\end{figure}

\textbf{Joule heating recipe.} Micromagnetic simulations of skyrmion
writing follow the Joule heating simulation recipe shown in \ref{Fig:S1_Sims-JHtg-SkWrite}a.
The system was initialized to at positive saturation (Step A) and
lowered to the base field (90\textbf{~}mT for \ref{Fig:S1_Sims-JHtg-SkWrite})
at 300~K while retaining the uniformly magnetized state, and allowed
to equilibrate for 5~ns (Step B). The temperature was then elevated
by $\Delta T$ for 30~ns (Step C), corresponding to the pulse duration.
During this time, the micromagnetic parameters, $M_{{\rm S}}$, $D$,
\emph{$K_{{\rm u}}$}, and \emph{$A$} were rescaled to values corresponding
to $T+\Delta T$ following scaling relations detailed in the main
text. Subsequently, the temperature was reset to 300~K, and the magnetic
parameters were restored to their RT values (Step D). The system was
then allowed to equilibrate at RT for 15~ns to produce the final
state (Step E).

\textbf{Nucleation threshold.} First, while the final state for \ref{Fig:S1_Sims-JHtg-SkWrite}b
($\Delta T$ = 111~K) remains uniformly magnetized, \ref{Fig:S1_Sims-JHtg-SkWrite}c
($\Delta T$ = 165~K) shows nucleated skyrmions. This supports the
existence of a threshold value of $\Delta T$ (between 111-165~K)
for skyrmion nucleation via Joule heating. Second, as seen in \ref{Fig:S1_Sims-JHtg-SkWrite}d,
no skyrmions are nucleated when the $\Delta T$ = 165~K simulations
are performed without rescaling magnetic parameters. This suggests
that the rescaling of magnetic parameters, which likely lowers the
energy barrier, is critical to this skyrmion nucleation mechanism.

\textbf{Nucleation snapshots.} \ref{Fig:S1_Sims-JHtg-SkWrite}c shows
that skyrmions are nucleated at elevated temperatures for $\Delta T$
= 165~K (Step C). When $\Delta T$ is removed, the configuration
remains largely the same (Step D), and further equilibration affects
a small fraction of the textures (Step E). However, this is not the
case in \ref{Fig:S1_Sims-JHtg-SkWrite}e, wherein $\Delta T$ is increased
to 228~K. In this case, large fluctuations at elevated temperatures
produce a randomly magnetized configuration (Step C). The removal
of $\Delta T$ may arbitrarily create a few nucleation (Step D), which
equilibrate to form skyrmions (Step E). While both \ref{Fig:S1_Sims-JHtg-SkWrite}c
and \ref{Fig:S1_Sims-JHtg-SkWrite}e produce skyrmion configurations,
their distinction is crucial. In case of \ref{Fig:S1_Sims-JHtg-SkWrite}e
parameters, successive current pulses should correspond to drastically
different magnetic configurations. In contrast, they may produce only
incremental changes for \ref{Fig:S1_Sims-JHtg-SkWrite}b parameters.

\section{Simulating Oersted Field Effects\label{sec:Sims_OeField}}

\begin{figure}[h]
\centering \includegraphics[width=0.6\textwidth]{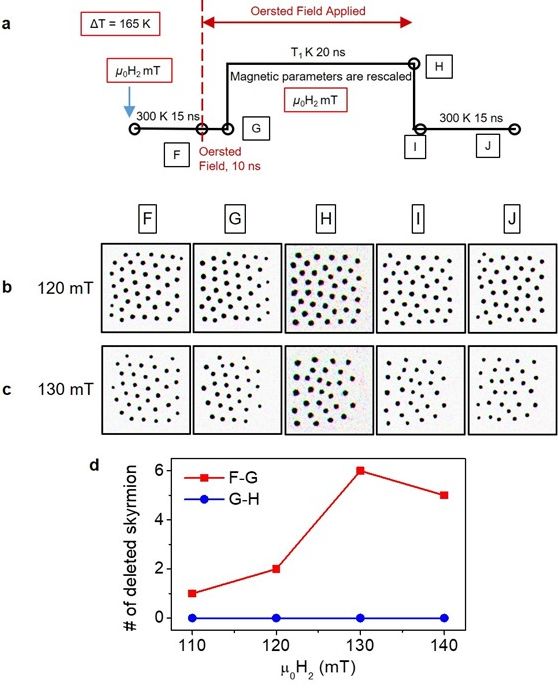}
\caption{\textbf{Simulations of Oersted field-driven skyrmion deletion.} \textbf{(a)}
Schematic of simulation recipe used to simulate Oersted field effects
at base field $\mu_{0}H_{2}$ with a temperature rise $\Delta T=$165~K
(as in \ref{Fig:S1_Sims-JHtg-SkWrite}. Labels F to J identify various
instances within the recipe. The Oersted field is applied for the
entire 30~ns pulse duration (starting at F), while Joule heating
is introduced for the final 20~ns (starting at G). \textbf{(b}-\textbf{c)}
Grayscale snapshots of simulated magnetization at instances F to J
{[}defined in (a){]} for (b) $\mu_{0}H_{2}=$ 120 mT and (c) $\mu_{0}H_{2}=$
130 mT, showing fewer skyrmions in J (c.f. F).}
\label{Fig:S2_Sims-OeField-SkDel}
\end{figure}

\textbf{Oersted Field Simulation Recipe.} Micromagnetic simulations
of skyrmion deletion follow the Oersted field effect simulation recipe
shown in \ref{Fig:S2_Sims-OeField-SkDel}a. The system is initialized
with the skyrmion configuration created in \ref{Fig:S1_Sims-JHtg-SkWrite}a
Step E. The field is raised to $\mu_{0}H_{2}$, and allowed to equilibrate
for 15~ns (Step F). At this point, the application of a current pulse
results in an instantaneous Oersted field and a delayed Joule heating
effect. These are simulated by applying an Oersted field with profile
in \textbf{Figure 3c} for the full current pulse duration and elevating
the temperature by $\Delta T$ {[}as in \ref{Fig:S1_Sims-JHtg-SkWrite}a{]}
after the pulse has been applied for 10~ns. This corresponds first
to a 10~ns time window with an Oersted field at 300~K (Step G),
and then to a 20~ns time window where the Oersted field and elevated
temperature are simultaneously in effect (Step H). Subsequently both
these effects are turned off as the system is restored to RT (Step
I), followed by equilibration for 15~ns to give the final state (Step
J).

\textbf{Annihilation Snapshots.} Skyrmion annihilation is seen for
both \ref{Fig:S2_Sims-OeField-SkDel}b ($\mu_{0}H_{2}=120$~mT) and
\ref{Fig:S2_Sims-OeField-SkDel}c ($\mu_{0}H_{2}=130$~mT): the final
state (J) has fewer skyrmions than the initial state (F). As clarified
in the main text, the annihilation is spatially asymmetric: skyrmions
are mostly annihilated on the right side of the image. Here, we clarify
the annihilation window by examining the magnetization snapshots across
the simulated duration. From inspection of \ref{Fig:S2_Sims-OeField-SkDel}b
and c, it is evident that the annihilation occurs predominantly in
the 10~ns (between F and G) wherein the Oersted field is present
at RT. In contrast, the comparison of snapshots after G, wherein the
temperature is also elevated, reveal negligible annihilation, albeit
with some random motion of skyrmions. Above statement can be more
clearly seen in \ref{Fig:S2_Sims-OeField-SkDel}d. we have tracked
the number of skyrmions at each step of the skyrmion deletion simulation
at different fields. \ref{Fig:S2_Sims-OeField-SkDel}d shows the number
of deleted skyrmions during the Oersted field only phase (F-G) and
the heating phase (G-H). It is evident that, during the heating phase
(G-H), there is no decrease in the number of skyrmions, and the decrease
in the number of skyrmions only takes place during the Oersted field
phase (F-G). This suggests that Joule heating effects counteract Oersted
field deletion by stabilizing skyrmions.

\begin{figure}[h]
\centering \includegraphics[width=0.6\textwidth]{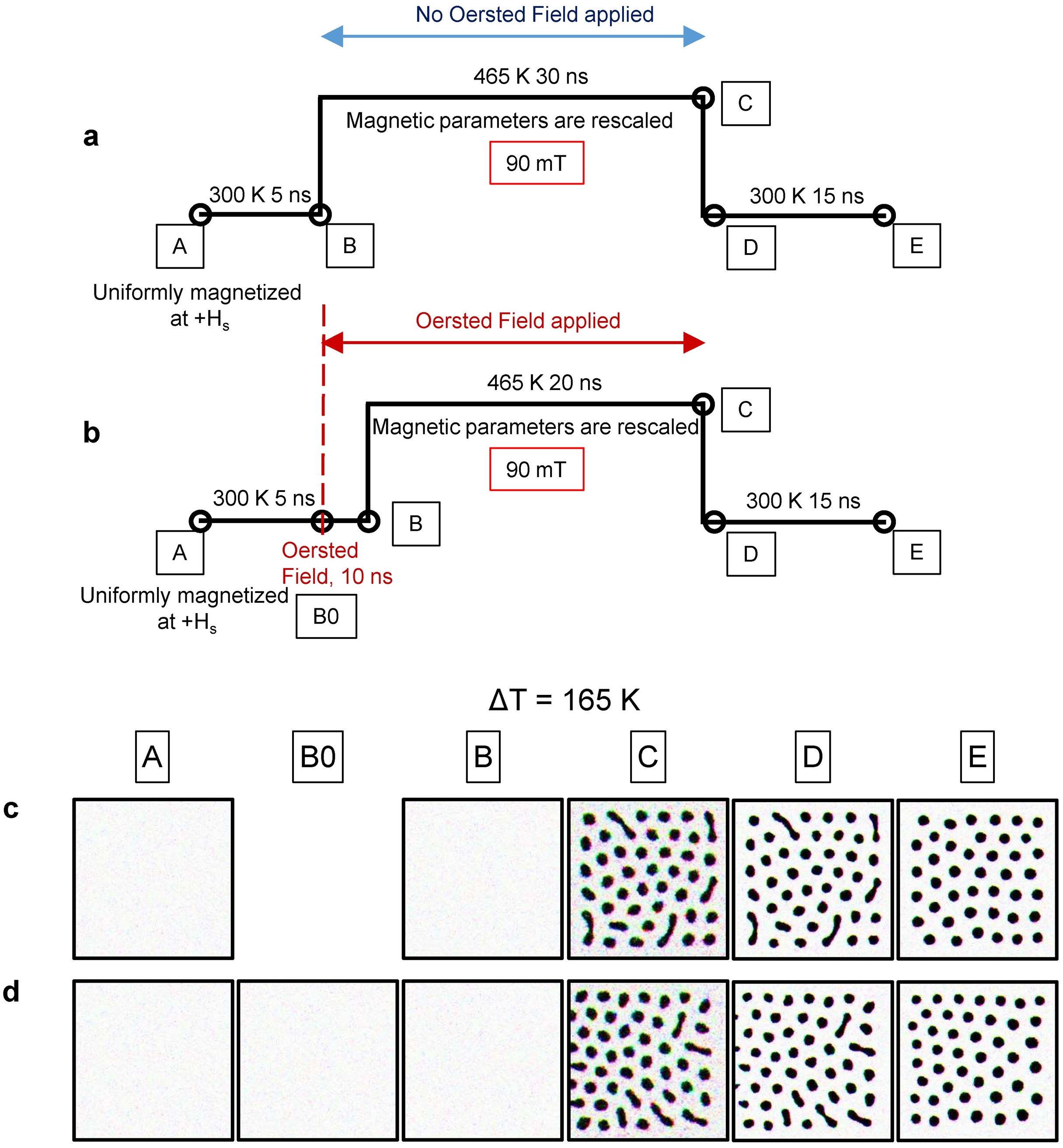}
\caption{\textbf{Simulations of Oersted field effects on skyrmion nucleation.}
\textbf{(a)} Simulation recipes for Joule heating-driven skyrmion
nucleation starting from a uniform state without Oersted fields. \textbf{(b)}
Simulation recipes for Joule heating-driven skyrmion nucleation with
Oersted fields. Labels A to E identify various instances within the
two recipes. \textbf{(c}-\textbf{d)} Grayscale snapshots of simulated
magnetization at instances A to E {[}defined in (a){]} for (c) no
Oersted field, and (d) with Oersted field.}
\label{Fig:S3_Sims-OeField-SkNuc}
\end{figure}

We note that Joule heating may have been associated with skyrmion
erasure in some previous works. In our case, through systematic simulations
detailed in \ref{Fig:S1_Sims-JHtg-SkWrite}, we have found that, over
a reasonably wide temperature ranges, the rescaling of micromagnetic
parameters associated with heating stabilizes existing skyrmions,
and in fact prefers to nucleate additional skyrmions. This allowed
us to conclude that the Oersted field is the main factor for the skyrmion
annihilation in our case. In this light, we conclude that works reporting
heat-based erasure of skyrmions may have explored a rather different
phase space regime.

\textbf{Oersted fields \& nucleation.} As the nucleation and annihilation
pulses are identical, their effects, i.e. Joule heating and Oersted
field, should appear in both cases. Having considered them in conjunction
during annihilation, we now revert to verifying if the introduction
of Oersted fields affects nucleation. In particular, we compare the
results from the \ref{Fig:S1_Sims-JHtg-SkWrite}a recipe (no Oersted
field) with those from the \ref{Fig:S2_Sims-OeField-SkDel}a recipe
(with Oersted field) for a uniformly magnetized initial state with
$\Delta T$ = 165~K (see \ref{Fig:S3_Sims-OeField-SkNuc}). Notably
we find little difference between the two sets of results shown in
\ref{Fig:S3_Sims-OeField-SkNuc}b-c in the intermediate or final stages.
Both produce skyrmion configurations with similar characteristics.
The diminished contribution of Oersted field effect for this case
is due to the uniformly magnetized state in the absence of Joule heating
effects (before B). This suggests that once skyrmions are created
at elevated temperatures (before C), their deletion is prevented by
the concomitant role of Joule heating. In summary, this validates
the simulation protocol followed in \ref{Fig:S1_Sims-JHtg-SkWrite}
(and Fig. 2 in the main text).

\section{Pinning Effects on Skyrmion Nucleation\label{sec:PinningEffects}}

\begin{figure}[h]
\centering \includegraphics[width=0.35\textwidth]{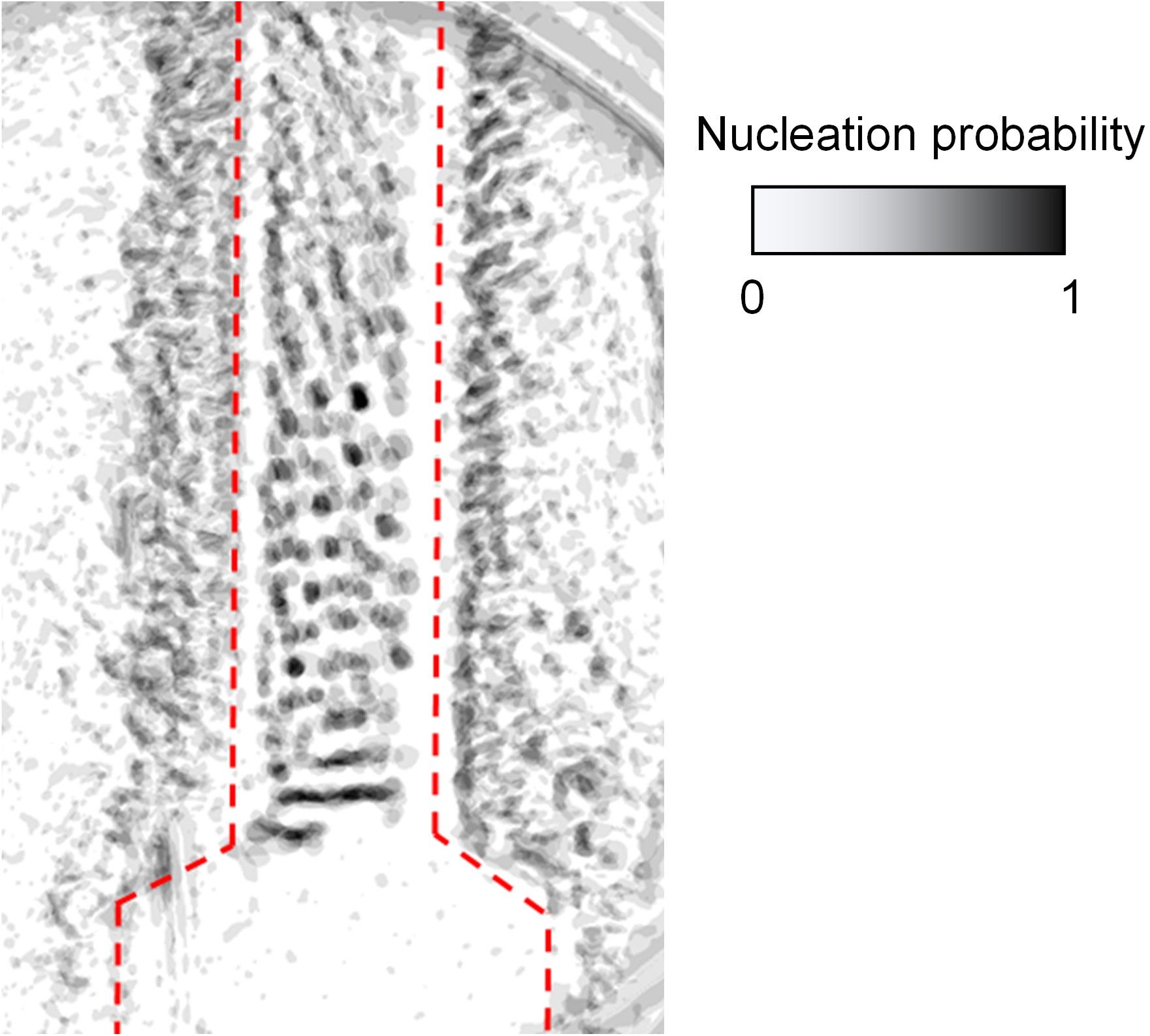}
\caption{Spatial distribution of the observed skyrmion nucleation obtained
by overlaying 11 MTXM images recorded after applying current pulses
to the uniformly magnetized wire device. Red dashed lines indicate
the wire edges. The contrast outside of the wire originates from irregularities
in the transparent membrane.}
\label{Fig:S4_MTXM-SkNuc_Overlaid}
\end{figure}

Within multilayer films, we expect that pinning effects may arise
from grain boundaries. These should lead to higher resistivity and
therefore to greater heating effects for the same current density.
However, from the Mayadas--Shatzkes model {[}Mayadas et al. Phys.
Rev. B 1. 1382 (1970){]}, electron scattering from grain boundaries
is only one of the possible resistive mechanisms. In ultrathin layered
films ($\sim1$~nm per layer), the dominant mechanism might be interface
scattering {[}Misra et al. J. Appl. Phys. 85, 302 (1999){]}. If this
is the case, the correlation between pinning and heating effects will
be weak. In order to experimentally address this question, one may
need to controllably vary the grain size, e.g., by tuning deposition
pressure or performing post-annealing, and subsequently measure the
residual resistivity ratio (RRR) of the resulting films. Such experiments
are beyond the scope of this manuscript, but nevertheless are important
directions for follow-up work.

\begin{figure}
\centering \includegraphics[width=0.5\textwidth]{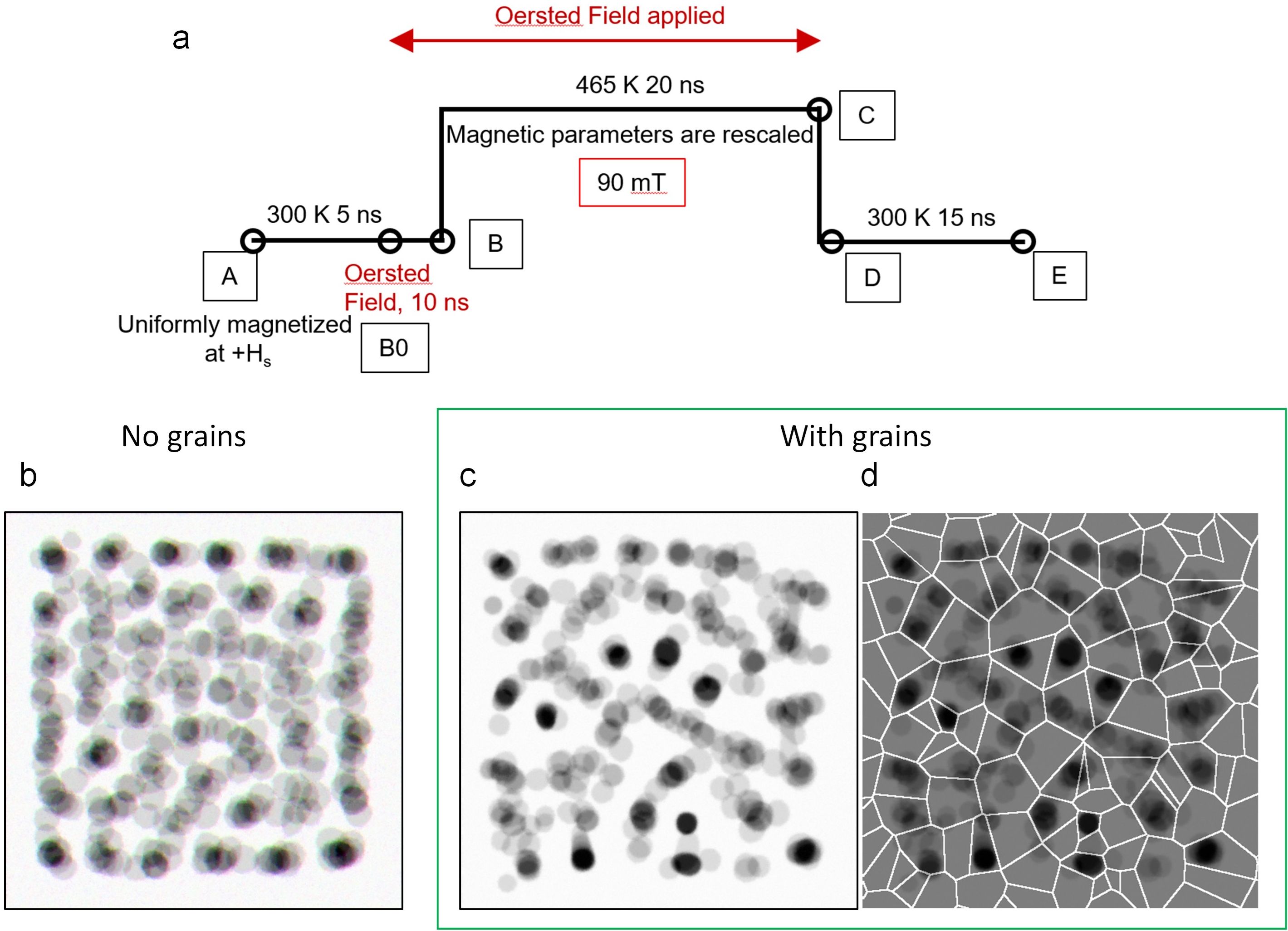}
\caption{\textbf{Micromagnetic simulation for the presence of grains.} Simulation
recipe for the skyrmion nucleation (a) and overlaid results from 8
simulations without grains (b) and with grains (c), with grain boundaries
shown in (d).}
\label{Fig:S5_Sims_SkNuc-Grains}
\end{figure}

Notwithstanding their contribution to heating, the question of whether
pinning effects are indeed occurring is equally relevant. To address
this, we have investigated the spatial distribution of skyrmion nucleation
probability in our device for both experimental and simulations. \ref{Fig:S4_MTXM-SkNuc_Overlaid}
shows the result obtained by overlapping 11 skyrmion nucleation images.
There are several darker spots, corresponding to a high skyrmion nucleation
probability. This indicates that our nucleation scheme depends on
the local variation of material characteristics, such as defects,
which could also serve as pining sites. However, at the moment, we
cannot confirm whether the high nucleation probability sites correspond
to pinning sites, as this would require investigating skyrmion motion
and the distribution of the pinning sites.

Meanwhile, to check if pinning effect could play a role in skyrmion
nucleation, we have performed additional simulations with the introduction
of random grains. We have compared the results from two sets of simulations
(repeated 8 times) with and without grains respectively. For the latter
case, the uniaxial anisotropy and interfacial DMI are varied by $10\%$
while all other magnetic parameters remain the same across grains
with average size $\sim200$~nm. \ref{Fig:S5_Sims_SkNuc-Grains}
shows the overlaid distributions of nucleated skyrmions obtained at
the final relaxed stage E in \ref{Fig:S5_Sims_SkNuc-Grains}a without
grains (b) and with grains (c). It is apparent that, with the introduction
of grains, skyrmion nucleation occurs more frequently in specific
grains, and the skyrmions are pinned at the edges of the grains (d).
This result probably explains the several darkest spots in \ref{Fig:S4_MTXM-SkNuc_Overlaid},
reflecting the presence of grains strongly affects the nucleation
positions of skyrmions.

\section{Skyrmion Size Analysis\label{sec:SkSize}}

\begin{figure}[h]
\centering \includegraphics[width=12cm]{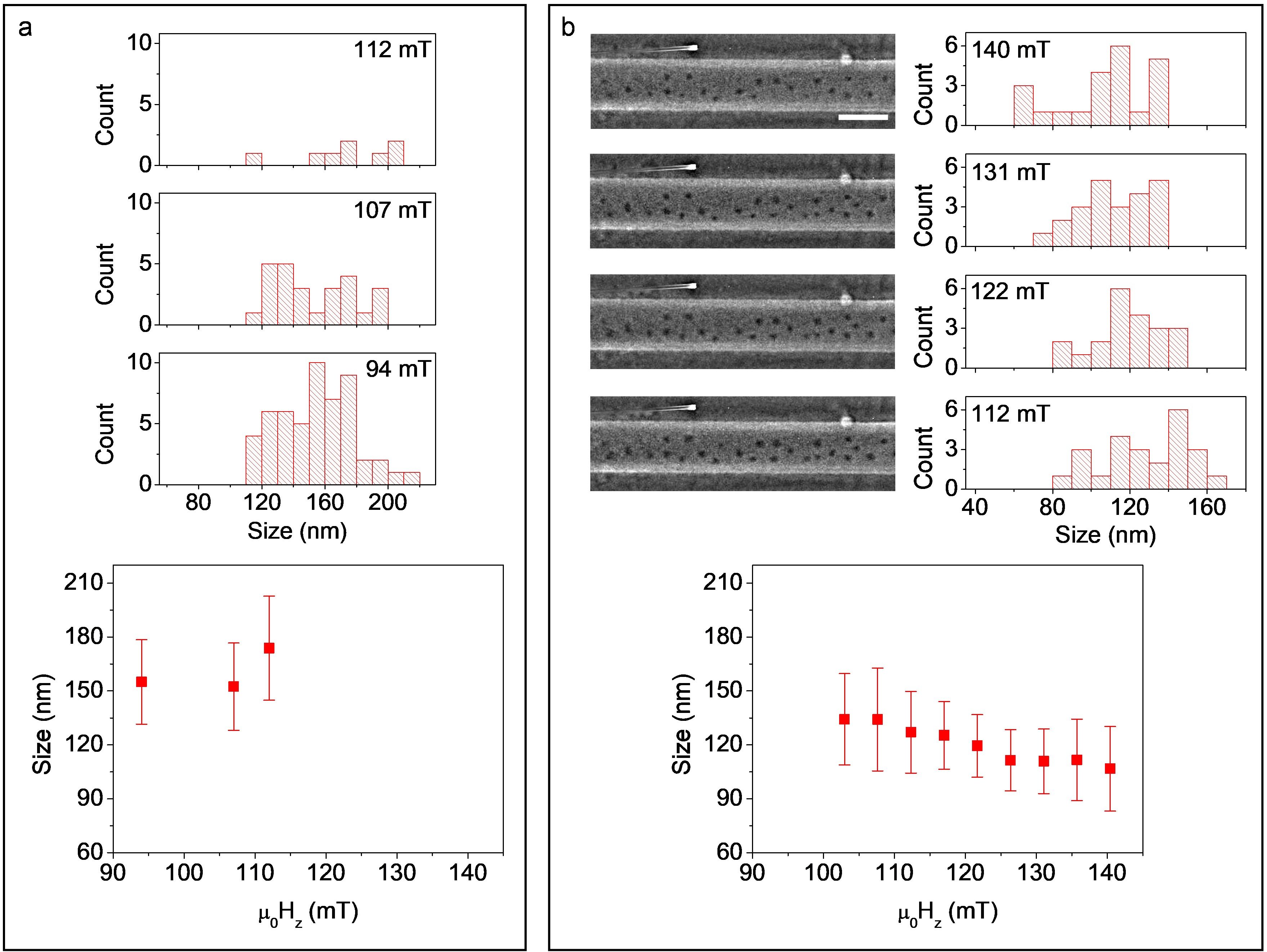}
\caption{\textbf{Analysis on skyrmion size with respect to the perpendicular
field.} (a) Analysis on the images from Figure 1d. (b) Analysis on
the results from the quasi-static experiments.}
\label{Fig:S6_SkSize}
\end{figure}

We performed analysis on the images in manuscript Fig. 1d for the
nucleated skyrmion size (\ref{Fig:S6_SkSize}a). Different from the
decreasing trend of the skyrmion size with increasing field, the data
does not show a decrease in the skyrmion size with the magnetic field.

We to note that the X-ray images in manuscript Fig. 1d are not optimized
to precisely determine skyrmion sizes, even though the data quality
is sufficient to determine the presence of skyrmions. We further caution
that the interpretation of size trends is constrained by limited nucleation
statistics at higher fields (10 times lower, see \ref{Fig:S6_SkSize}a
top). In summary, the nucleated skyrmion images does not have sufficient
statistical significance to make meaningful claims about skyrmion
size trends with magnetic field.

However, to clarify the evolution of skyrmion size with magnetic field,
we performed additional experiments with optimized microscopy conditions.
The results in \ref{Fig:S6_SkSize}b show a pronounced decrease in
skyrmion size with increasing magnetic fields in line with the simulation
result in manuscript Fig. 2c.

Please note that, in this additional experiment, initial skyrmions
were nucleated at the low field, and we tracked their size as increasing
the field. This quasi-static experimental protocol is different from
the experiments shown in manuscript Fig. 1d where skyrmions are distinctly
nucleated at each measured field. Therefore, we cannot completely
exclude the possibility that the sizes of the skyrmions from the two
cases cannot be matched. Notably, one could have an inconsistency
in skyrmion sizes if only bigger skyrmions survive the current pulse
induced thermal fluctuations once nucleated at higher field. However,
in the quasi-static case, small skyrmions may also survive in higher
fields because there are no significant thermal fluctuations. The
effect of pinning could also influence the two cases in different
ways.

\section{Calculation of the Oersted Field\label{sec:OeFieldCalc}}

\begin{figure}[h]
\centering \includegraphics[width=0.75\textwidth]{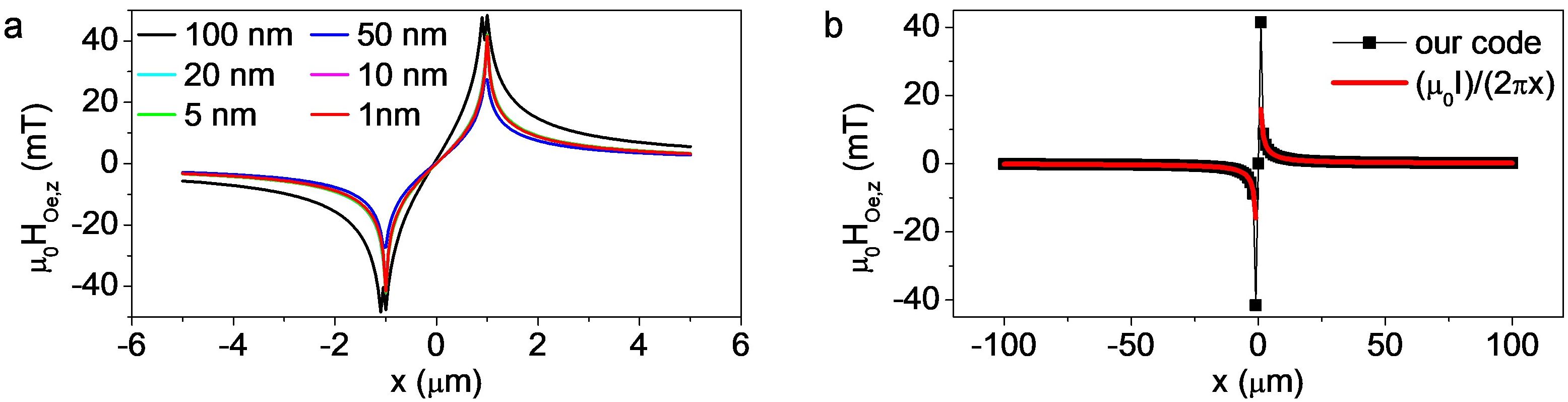}
\caption{\textbf{(a)} Perpendicular component of the Oersted field profile
calculated for various sizes of the wire segments and \textbf{(b)}
comparison between calculated Oersted field profile (perpendicular)
and that from the Ampère's circuital law for a straight line.}
\label{Fig:S7_OeFieldCalc}
\end{figure}

The Oersted field in the manuscript was calculated by superposing
the magnetic fields generated from wire segments (1~nm$^{2}$ cross-section)
through the wire cross-section (2~$\mu$m $\times$ 58~nm). Uniform
current density was assumed and the Biot-Savart Law was used. The
wire segments of 1~nm cell were chosen because the cell-size dependence
of calculated magnetic field profile disappears as the cell-size gets
below 10 nm as shown in \ref{Fig:S7_OeFieldCalc}a. To test the validity
of our code, the calculated magnetic field profile is compared with
the result of the well-known equation $B=(\mu_{0}/2\pi)\cdot I\,/\,x$,
the magnetic field from a straight line (Ampère's circuital law),
as shown in \ref{Fig:S7_OeFieldCalc}b. It shows perfect agreement
with the equation outside of the wire ($|x|>1$~$\mu$m), confirming
that our code for the Oersted field calculation is correct.

\section{Skyrmion Annihilation for Reversed Field and Current \label{sec:Sk-DeletionCheck}}

\begin{figure}[h]
\centering \includegraphics[width=0.6\textwidth]{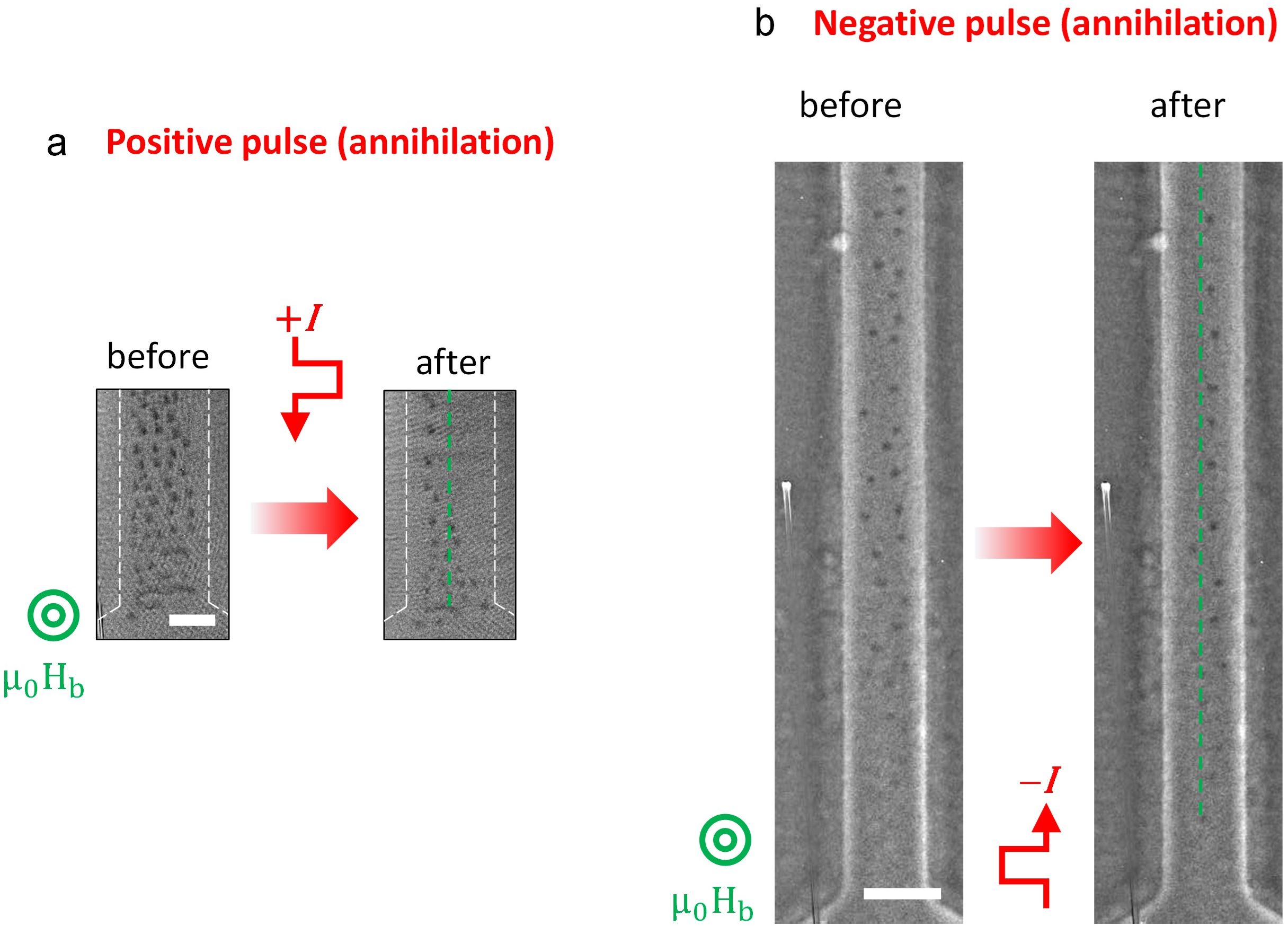}
\caption{Skyrmion annihilation by the application of a \textbf{(a) }positive
current pulse and \textbf{(b)} negative current pulse (scale bar:
1 $\mu$m).}
\label{Fig:S8_SkDelCheck}
\end{figure}

To check if the spatial asymmetry of the skyrmion distribution in
the annihilation experiment indeed originates from the Oersted field,
we performed skyrmion annihilation experiments with reversed current
pulse. The result is displayed in \ref{Fig:S8_SkDelCheck}b and is
obtained from a different wire but on the same film. The images in
\ref{Fig:S8_SkDelCheck}a are taken from manuscript Fig. 3a for comparison.
It is evident that the skyrmion annihilation is dominant on the left
side of the wire for the negative pules (\ref{Fig:S8_SkDelCheck}b)
while the annihilation is dominant on the right side of the wire for
the positive current pules (\ref{Fig:S8_SkDelCheck}a). These results
are in perfect agreement with the presence Oersted field, confirming
the Oersted field effect.
\begin{center}
{\small{}\rule[0.5ex]{0.6\columnwidth}{0.5pt}}{\small\par}
\par\end{center}

\end{document}